\documentclass[fleqn,usenatbib]{mnras}
\newcommand*\Bell{\ensuremath{\boldsymbol\ell}}
\usepackage{float}

\usepackage{newtxtext,newtxmath,xcolor}
\usepackage{pdflscape}

\usepackage[T1]{fontenc}
\usepackage{siunitx}
\DeclareRobustCommand{\VAN}[3]{#2}
\let\VANthebibliography\thebibliography
\def\thebibliography{\DeclareRobustCommand{\VAN}[3]{##3}\VANthebibliography}

\usepackage{graphicx}	
\usepackage{amsmath}	
\usepackage{float}

\title[Dynamical dust traps]{Dynamical dust traps in misaligned circumbinary discs: analytical theory and numerical simulations}

\author[C. Longarini et al.]{
Cristiano Longarini,$^{1}$\thanks{E-mail: cristiano.longarini@unimi.it}
Giuseppe Lodato,$^{1}$
Claudia Toci$^{1}$
and Hossam Aly$^{2}$
\\
$^{1}$Dipartimento di Fisica, Università degli Studi di Milano, via Celoria 16, 20133 Milano, Italy\\
$^{2}$Univ Lyon, Univ Claude Bernard Lyon 1, Ens de Lyon, CNRS, Centre de Recherche Astrophysique de Lyon UMR5574, F-69230, Saint-Genis-Laval, France
}

\date{Accepted XXX. Received YYY; in original form ZZZ}

\pubyear{2021}

\begin{document}
\label{firstpage}
\pagerange{\pageref{firstpage}--\pageref{lastpage}}
\maketitle

\begin{abstract}
Recent observations have shown that circumbinary discs can be misaligned with respect to the binary orbital plane.The lack of spherical symmetry, together with the non-planar geometry of these systems, causes differential precession which might induce the propagation of warps. While gas dynamics in such environments is well understood, little is known about dusty discs. In this work, we analytically study the problem of dust traps formation in misaligned circumbinary discs. We find that pile-ups may be induced not by pressure maxima, as the usual dust traps, but by a difference in precession rates between the gas and dust. Indeed, this difference makes the radial drift inefficient in two locations, leading to the formation of two dust rings whose position depends on the system parameters. This phenomenon is likely to occur to marginally coupled dust particles $(\text{St}\gtrsim1)$ as both the effect of gravitational and drag force are considerable. We then perform a suite of three-dimensional SPH numerical simulations to compare the results with our theoretical predictions. We explore the parameter space, varying stellar mass ratio, disc thickness, radial extension, and we find a general agreement with the analytical expectations. Such dust pile-up prevents radial drift, fosters dust growth and may thus promote the planet formation in circumbinary discs.
\end{abstract}

\begin{keywords}
hydrodynamics - accretion, accretion discs - protoplanetary discs - planets and satellites: formation - methods: analytical - methods: numerical
\end{keywords}



\section{Introduction}
Most stars in our Galaxy form in clustered environments \citep{Clarkecluster}. It is then reasonable to expect that pre-stellar objects dynamically interact with each other, eventually forming multiple systems \citep{bate2009}. In particular, around young binary systems three discs can be found: two around the stars (circumprimary and circumsecondary) and a larger circumbinary disc. Surveys of star forming environments \citep{mckee}, as well as numerical studies  \citep{bate2018} show that stars form from a sequence of accretion episodes, for which the angular momentum can be randomly oriented. In such a scenario, the disc morphology hardly remains flat: at least some circumbinary discs are expected to be misaligned to the binary orbital plane or warped. Recent observations have confirmed these expectations, for example in the sources HD 98800B \citep{czekala}, GG Tau A \citep{GGTAUA1, GGTAUA2}, KH 15D \citep{chiang,Lodato2014,pascucci}. In addition, circumbinary discs are  environments where planetary formation takes place: this has been  indirectly verified by means of timing of stellar eclipse \citep{deeg,lee,beuermann} and directly confirmed in 2011 \citep{doyle} from Kepler spacecraft data. While the gas dynamics in non-planar and warped disc has been deeply studied \citep{warp1,warp2,ogilvie99, Facchini2014}, dust dynamics has not yet been thoroughly investigated analytically, but only by means of numerical simulations \citep{AlyLod}.

In protoplanetary discs, gas and dust are aerodynamically coupled by means of a drag force caused by the different velocity of the two components. The gas has a sub-keplerian azimuthal velocity due to its pressure gradient and a radial velocity given by viscous effects. On the contrary, the dust velocity is almost keplerian, being a pressureless and viscousless  fluid. Thus, the resulting drag force tends to slow the dust down due to the azimuthal headwind and lets solid particles migrate toward the central object. There are two different drag force regimes, depending on the solid particles' size. For typical protoplanetary discs and particle size less than 1 meter, the gas mean free path is much larger than the dust particle size $(\lambda_g>s)$, hence the drag force is well described by the Epstein regime \citep{epstein}
\begin{equation}
    \mathbf{F}_\text{D}^{\text{Epst}} = -\frac{4 \pi}{3} \rho_{\mathrm{g}} s^{2} u_{\mathrm{th}} \Delta \mathbf{u},
\end{equation}
where $\rho_g$ is the gas density, $s$ the particles' dimension, $u_\text{th}=\sqrt{8 / \pi} c_{s}$ is the mean thermal velocity and $\Delta\mathbf{u}=\mathbf{u}_d-\mathbf{u}_g$ with $\mathbf{u}_d$ the dust velocity and $\mathbf{u}_g$ the gas one. A measure of the strength of this coupling is the stopping time, i.e. the time in which drag modifies the relative velocity significantly.
\begin{equation}
    t_{s}=\frac{m_{d}|\Delta \mathbf{u}|}{\left|\mathbf{F}_{D}\right|} \underset{\text{Epstein}}{=}\frac{\rho_{0} s}{\rho_{g} u_\text{th}},
\end{equation}
where $\rho_0$ is the intrinsic grains' density. However, a more useful quantity to describe the coupling is by considering the Stokes number, i.e. the ratio between the stopping time and the dynamical one
\begin{equation}
    \text{St} = \frac{t_s}{t_\text{dyn}} = t_s\Omega_k \underset{\text{Epstein}}{=} \frac{\pi}{2} \frac{\rho_{0} s}{\Sigma_{g}},
\end{equation}
where $\Sigma_g = \rho_g H = \rho_g c_s /\Omega_k$ is the gas surface density. Solid particles with $\text{St}<<1$ are strongly coupled to the gas, while the ones with $\text{St}>>1$ are weakly coupled.
This has important implications on planet formation: while small  particles follow the dynamics of the gas, large sized grains  drift inwards within a small fraction of the disc lifetime, preventing planetesimals formation.

The evolution of the dust is crucial for the standard theory of planet formation. According to the ``core accretion'' theory \citep{safronov, Goldreich}, solid particles, growing from micron-size to kilometre-size, form planetary cores and when they reach enough mass, they start accreting gas.  However, this process is hindered by the radial drift of solid particles due to their aerodynamical coupling with the gas.  The effect of radial drift is maximum for particles with $\text{St}=1$ \citep{armitage}, corresponding to particle size of cm-m: this is the so called ``metre sized barrier" \citep{metresize} to planet formation.

A possible way to prevent radial drift and overcome the metre sized barrier is the presence of a ``dust trap", i.e. a mechanism that piles up dust particles in gas local pressure maxima \citep{nakagawa}. Indeed, when the pressure gradient is zero, the difference of speed between gas and dust vanishes and no radial drift occurs. Several mechanisms have been proposed to form dust traps, such as gaps made by planets \citep{Pinilla}, magnetic winds \citep{magn}, spirals induced by gravitational instabilities \citep{spiralarms} or self-induced dust traps \citep{selfinduced1,selfinduced2}.

In this paper, we study a mechanism that leads to the formation of dust rings in misaligned circumbinary discs. This phenomenon has been numerically found by \citet{AlyLod}; here we give an analytical explanation and we present more simulations of this phenomenon. The lack of spherical symmetry in the gravitational potential with the non-planar geometry of the system makes gas and dust precess at different rates: the gaseous disc precesses as a rigid body while the dusty disc precesses differentially. In a system like this, three significant radii can be found, the co-precession radius, the inner dusty radius, and the outer dusty radius. At the co-precession radius ($R_\text{cp}$), the two precession velocities are equal, while at the inner and outer dusty radii ($R_\text{IDR},R_\text{ODR}$) the velocity difference, including pressure corrections, is exactly zero and two dust rings form. At such radii, solid particles, migrating from the outer to the inner part of the disc because of radial drift, stop, pile up and form rings. This mechanism acts as a dust trap because when the difference of speed between gas and dust is zero, the drag force does not act, and thus the radial drift is stopped. This phenomenon is deeply different from the classical dust traps because it does not require a pressure maximum in the gaseous component.

The paper is organised as follows. In section \ref{S2} we  present the analytical theory: we write the velocity fields of both components, we find three significant locations on the disc and we discuss the role of the aerodynamical coupling. In section \ref{S3}, we explain the numerical methods: we introduce the numerical code we have used and we describe the simulations we have performed. In section \ref{S4} we discuss the results, making a comparison between theoretical and numerical methods. In section \ref{S5} we connect our work to the observations, studying the observability of these systems. In section \ref{S6} we presents the conclusions of this work.

\section{Analytical theory}\label{S2}

\subsection{Velocity fields}\label{velocityfields}
A non-planar disc can be described in terms of two angles, the tilt and the twist angle. If we consider a ring of material at a distance $R$ from the central object, the angular momentum unit vector can be written as
\begin{equation}
    \Bell = (\cos\gamma\sin\beta,\sin\gamma\sin\beta,\cos\beta),
\end{equation}
where $\beta$ is the tilt and $\gamma$ the twist angle. A non-planar disc is warped if $\beta = \beta(R)$ and it is twisted when, additionally, $\gamma=\gamma(R)$. 

We consider two masses $M_1$ and $M_2$, gravitationally bound with a semi-major axis $a$, orbiting around their centre of mass. The second order, time independent, gravitational potential generated by the binary system has been computed by \citet{Facchini2014} and it is:
\begin{equation}
    \Phi(R, z)=-\frac{G M}{R}-\frac{G M \eta a^{2}}{4 R^{3}}+\frac{G M z^{2}}{2 R^{3}}+\frac{9}{8} \frac{G M \eta a^{2} z^{2}}{R^{5}},
\end{equation}
where $\eta = M_1M_2/(M_1+M_2)^2$ and $(R,z)$ are cylindrical coordinates with the origin in the centre of mass. We now consider a circumbinary disc, composed by gas and dust, with inner radius $R_\text{in}$ and outer radius $R_\text{out}$, misaligned to the binary plane with an angle $\beta$. Gas and dust dynamics are deeply different: while the gas tends to precess rigidly with a typical frequency $\omega_p$, the dusty disc precesses differentially with a frequency $\Omega_p(R)$, induced at a given radius $R$ by the binary. The dust differential precession rate is
\begin{equation}
    \Omega_{p}(R)=\frac{3}{4} \frac{\sqrt{G M} \eta a^{2}}{R^{7 / 2}},
\end{equation}
while the rigid precession of the gas, $\omega_p$, can be easily computed knowing the surface density profile of the gas $\Sigma$, as
\begin{equation}
    \omega_p = \frac{\int_{R_{\text {in }}}^{R_{\text {out }}} \Omega_{p }(R) L(R) 2 \pi R \mathrm{d} R}{\int_{R_{\text {in }}}^{R_{\text {out }}} L(R) 2 \pi R \mathrm{d} R} = \Omega_p(R_\text{in})\xi,
\end{equation}
where $L(R)=\Sigma R^2 \Omega$ and $\xi$ depends only on the ratio between the inner and the outer disc radii, $x_\text{out}= R_\text{in}/R_\text{out}$ and on the functional form of the surface density profile \citep{Lodato2014,align2}. For the complete derivation of these quantities see appendix \ref{appA}.

We now explicitly write the velocity fields of the two components to find the points of the disc where their difference is zero: in those point no radial drift occurs. To do this, we start from the velocity field of a flat disc in Carthesian coordinates and we rotate it by an angle $\beta$ around the $y-$axis and by an angle $\gamma$ around the $z-$axis. The geometry of the gas is described by $\beta_g =\beta$ and $\gamma_g=\omega_p t$, while the one of the dust by $\beta_d = \beta_g$ and $\gamma_d = \Omega_p(R)t$. In addition, we recall that the flat velocity of the gas is slightly sub-keplerian, while the dusty one is keplerian. Considering this, the velocity fields can be written as:
\begin{equation}
    \mathbf{u}_\text{g} = R\Omega_k\left(1+\frac{\zeta}{2}\right)\left[\begin{array}{c}-\sin \phi \cos (\omega_pt) \cos \beta-\cos \phi \sin (\omega_pt )\\ \cos \phi \cos (\omega_pt)-\sin \phi \cos \beta \sin (\omega_pt) \\ \sin \phi \sin \beta\end{array}\right],
\end{equation}
\begin{equation}
    \mathbf{u}_\text{d} = R\Omega_k\left[\begin{array}{c}-\sin \phi \cos (\Omega_p(R)t )\cos \beta-\cos \phi \sin (\Omega_p(R)t )\\ \cos \phi \cos (\Omega_p(R)t)-\sin \phi \cos \beta \sin( \Omega_p(R)t) \\ \sin \phi \sin \beta\end{array}\right],
\end{equation}
where, if we consider a density profile $\Sigma_g\propto R^{-p}$, the pressure correction is $\zeta = \text{d}\log P / \text{d} \log R\simeq -p(H/R)^2$. It is important to remark that in this analysis the radial velocity of the gas has been neglected because it is much smaller than the azimuthal one and we also have assumed the gas and dust to be co-planar $(\beta_g=\beta_d=\beta)$. It is well known that the viscosity tends to align the disc with the binary plane \citep{align1,align2,align3}, however we do not consider it in our analytical model. Further details about this phenomenon can be found in section \ref{alignmentsection}.

\subsection{Co-precession radius}
\begin{figure}
    \centering
    \includegraphics[scale=0.45]{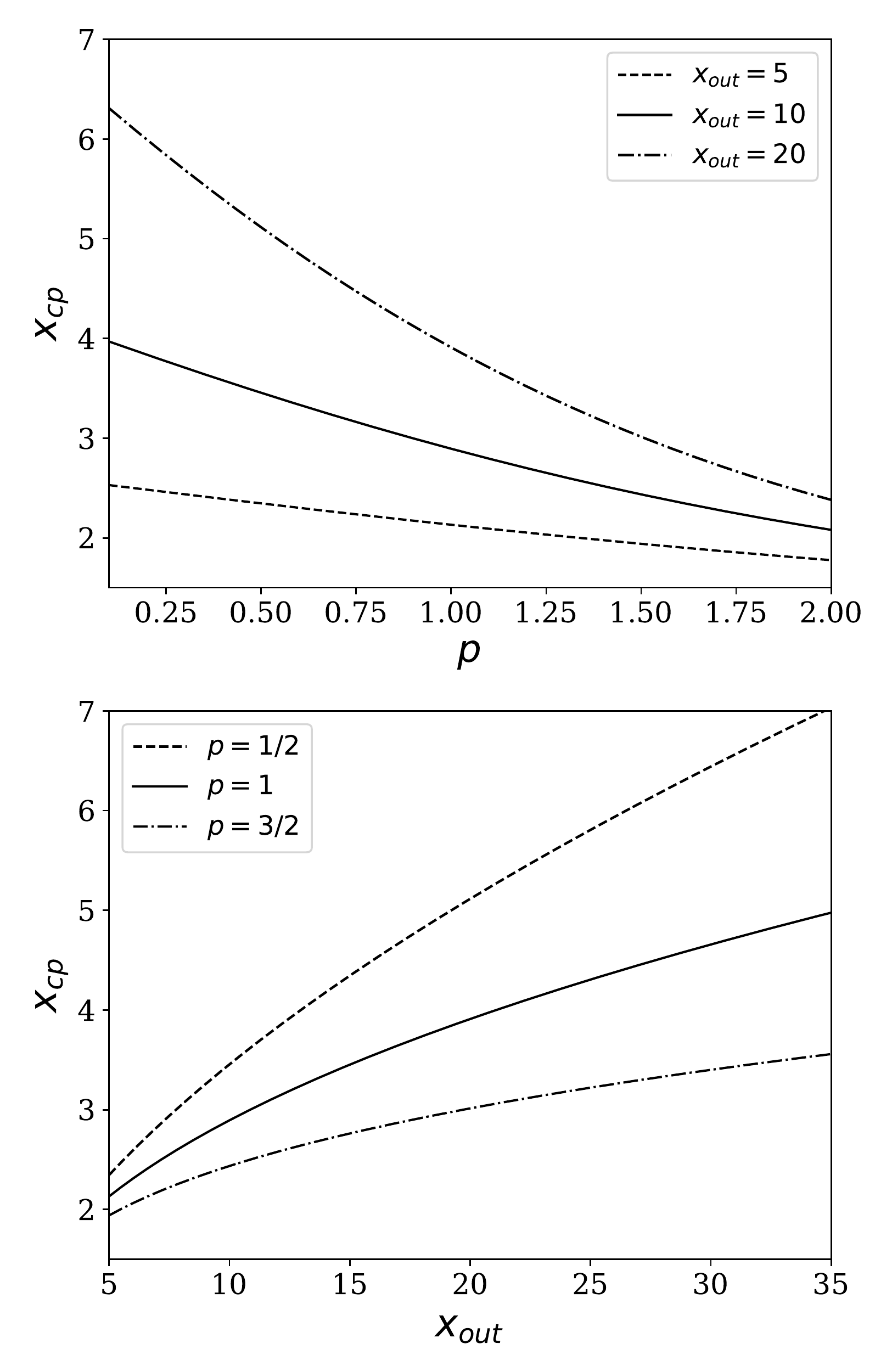}
    \caption{Adimensional co-precession radius $x_\text{cp}=R_\text{cp}/R_\text{in}$ as a function of the power law index of the density profile $p$ and of the adimensional outer radius $x_\text{out}=R_\text{out}/R_\text{in}$, for a disc with a pure power law density profile $\Sigma_g = \Sigma_0R^{-p}$.}
    \label{coprecessionfig}
\end{figure}
A significant location in the circumbinary disc is where the precession profiles of gas and dust meet, the co-precession radius. It can be easily computed solving $\Omega_p(R)=\omega_p$, that gives
\begin{equation}
    R_\text{cp} = R_\text{in} \xi^{-2/7}.
\end{equation}
Figure \ref{coprecessionfig} shows how the position of the co-precession radius varies as a function of the power law index of the density profile $p$, assuming a pure power law profile, and of the disc radial extension. At this radius, the two components precess at the same rate. However, due to the sub-keplerian gas speed, the aerodynamical coupling between gas and dust still produces a headwind: at this radius, dust particles still drift towards the central object. 

\subsection{Dust rings}
\begin{figure*}
    \centering
    \includegraphics[scale=0.4]{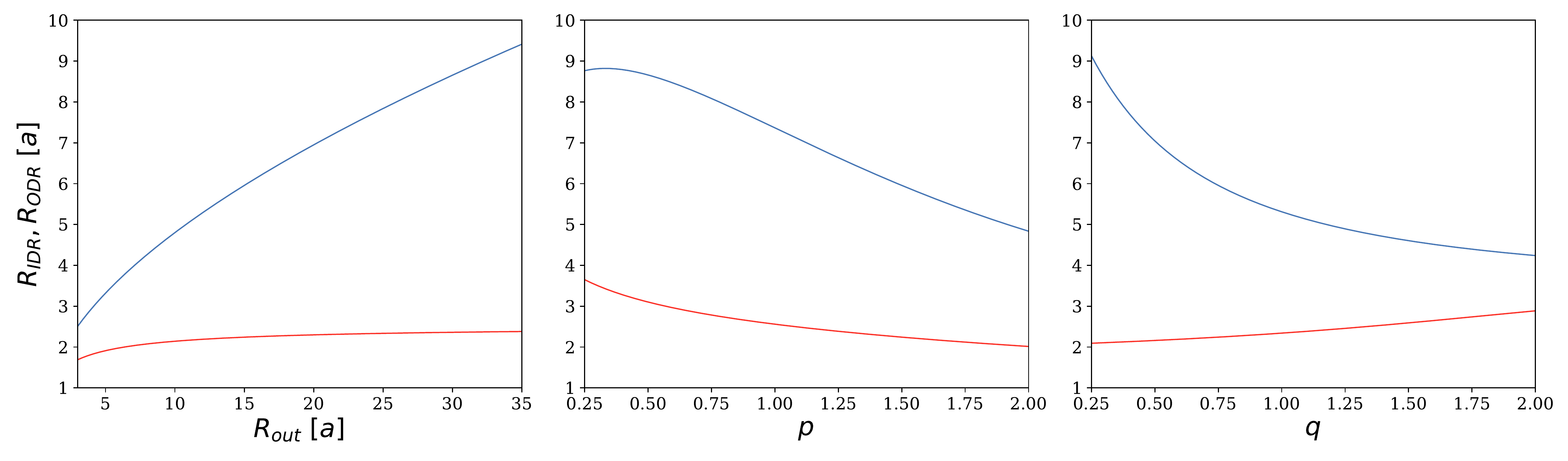}
    \includegraphics[scale=0.4]{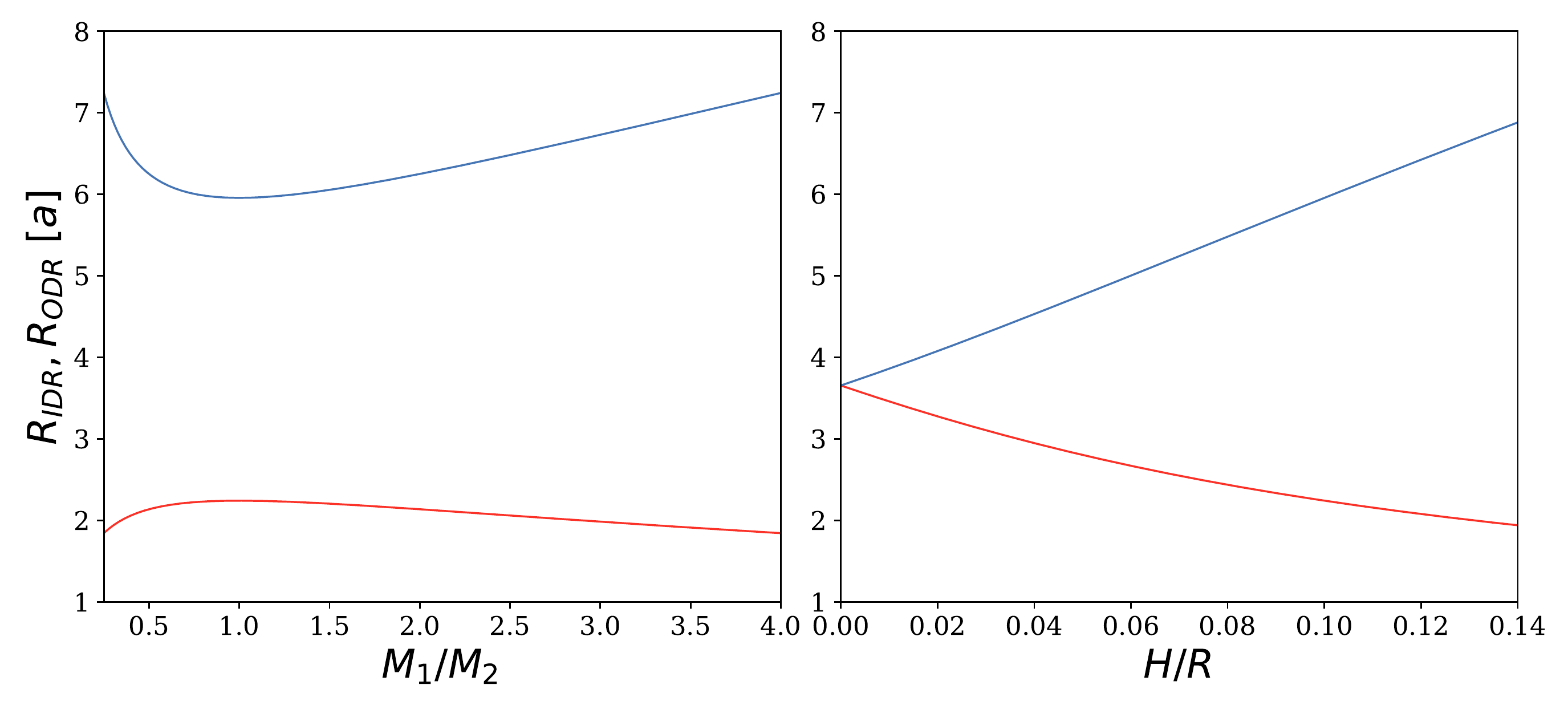}
    \caption{Position of dust rings as a function of the circumbinary disc extension R$_{\rm out}$, the power law index of the density profile $p$, the disc thickness $H/R$, the binary mass ratio M$_1$/M$_2$ and the power law index of the sound speed $q$. The standard parameters used in these plots are: $R_\text{in}=\SI{15}{AU}$, $R_\text{out}=\SI{150}{AU}$, $p=3/2$, $q=3/4$, $H/R=0.1$ and $M_1/M_2=1$ . }
    \label{position}
\end{figure*}
Here, we are interested in finding any locations in the disc where the total difference of speed -- including pressure gradients -- between gas and dust is exactly zero. Hence, we need equality between the dust velocity projected onto the gaseous plane (introduced in section \ref{velocityfields}) and the sub-keplerian speed of the gas. To properly project the velocity, we need to compute the angle between gas and dust at any radius $R$, obtained through the scalar product between their angular momentum unit vectors. We then find:
\begin{equation}
    \cos \psi=\Bell_{g} \cdot \Bell_{d}=1-\sin ^{2} \beta(1-\cos \Delta \omega t),
\end{equation}
where $\Delta\omega=\Omega_p-\omega_p$. To explicitly evaluate the radius where the projection of the dust velocity and the gaseous one are equal one has to solve the following equation: 
\begin{equation}
    \left|\mathbf{u}_{d}\right| \cos \psi=\left|\mathbf{u}_{g}\right| \rightarrow u_{k}\left[1-\sin ^{2} \beta(1-\cos \Delta \omega t)\right]=u_{k}\left[1+\frac{\zeta}{2}\right], 
\end{equation}
that, after some algebra, gives
\begin{equation}\label{proj}
    \cos \Delta \omega t=1-\frac{p\left(\frac{H}{R}\right)^{2}}{2 \sin ^{2} \beta}.
\end{equation}
In principle this expression depends on time ($t$) and radius ($\Omega_p= \Omega_p(R))$. However, in the system the precession timescale is bigger than the orbital one. We can then perform an orbital time average, obtaining an expression depending only on radius:
\begin{equation}
    \langle\cos \Delta \omega t\rangle=\frac{1}{2 \pi} \int_{0}^{2 \pi} \mathrm{d} \phi \cos \left(\frac{\Delta \omega}{\Omega_{k}}\right)=\frac{\Omega_{k}}{2 \pi \Delta \omega} \sin \left(\frac{2 \pi \Delta \omega}{\Omega_{k}}\right).
\end{equation}
Introducing the quantity $\varpi = {2 \pi \Delta \omega}/{\Omega_{k}}$, the previous condition (eq. \ref{proj}) reads
\begin{equation}\label{finalone}
    \frac{\sin \varpi}{\varpi}=1-\frac{p\left(\frac{H}{R}\right)^{2}}{2 \sin ^{2} \beta}.
\end{equation}
Equation (\ref{finalone}) has two solutions, $R_\text{IDR}$ and $R_\text{ODR}$, the position of an inner and an outer dust ring. These solutions depends on all the parameters of the disc, such as the size of the disc R$_{\rm out}$ and on the mass ratio of the stars, that determines the differential precession rate. Figure \ref{position} shows the positions of the dust rings as a function of the relevant parameters. 

Note that this mechanism influences only the dust density profile, leaving the gaseous one unperturbed. This differs from conventional dust traps mechanisms because the dust rings do not correspond to any maximum in the pressure profile of the gas.

\subsection{The role of the aerodynamical coupling}
So far, we have neglected the effect of the drag force on the dust precession profile: this is correct if the coupling between gas and dust is negligible ($\text{St}>>1$), however, when the Stokes number is smaller, this analysis has to be reviewed. 

\begin{figure}
\hspace{0.75cm}
    \includegraphics[scale=0.455]{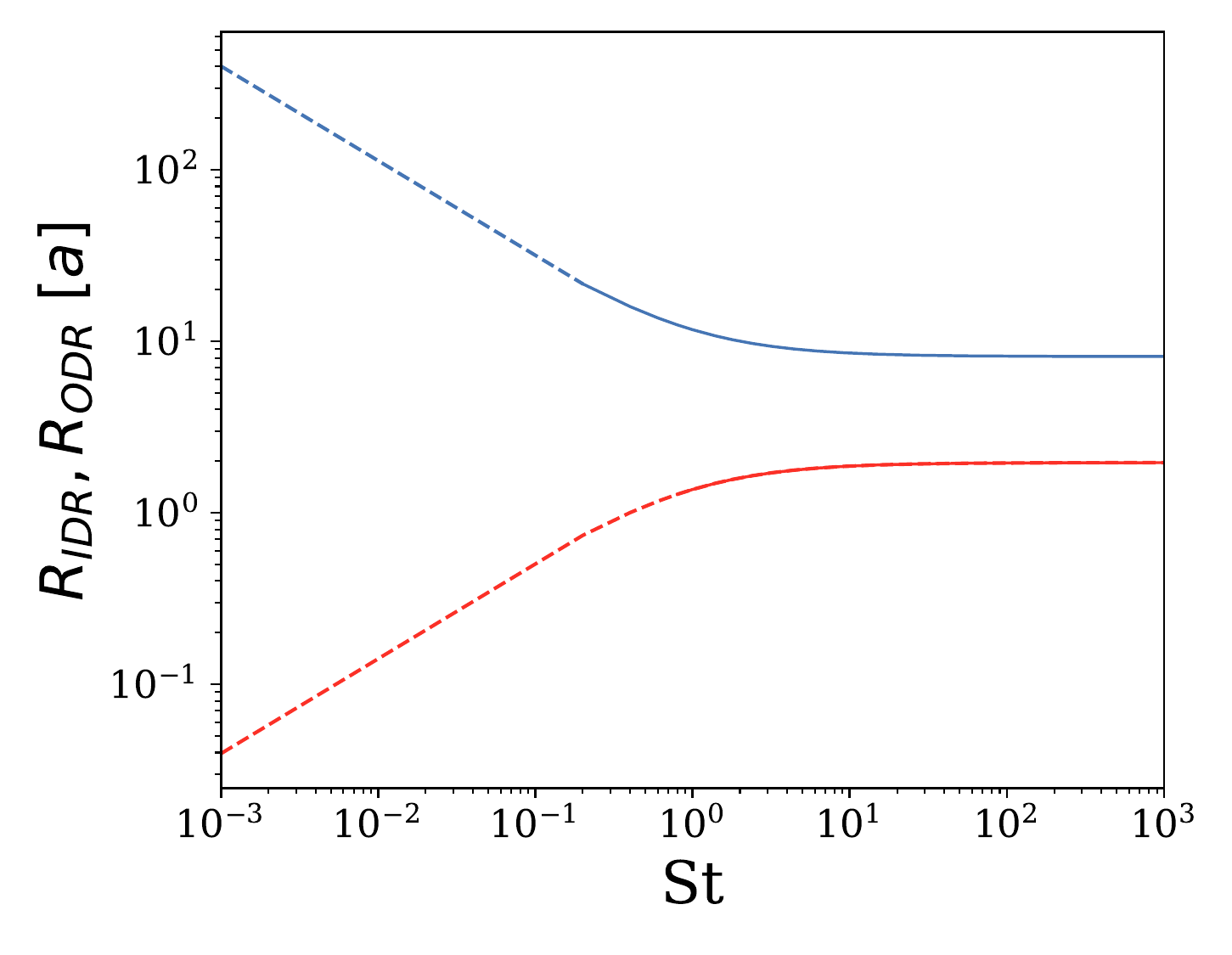}
    \caption{Position of dust rings as a function of the Stokes number. The standard parameters used in these plots are: $R_\text{in}=\SI{15}{AU}$, $R_\text{out}=\SI{150}{AU}$, $p=3/2$, $H/R=0.1$, $M_1/M_2=1$ and $q=3/4$. The solid line means that the position of the dust rings is inside the disc, while the dashed line means that the position is outside the disc, so no dust rings occur.}\label{ringsStokes}
\end{figure}

For weakly coupled dust particles , we recall that $\gamma_{d} =\Omega_{p}(R) t$ and $\gamma_{g}=\omega_{p} t$. On the other hand, we know that for strongly coupled dust particles ($\text{St} <<1$ ), the dust precession profile corresponds to the gaseous one, meaning that $\gamma_{d}(\text{St}<<1)=\gamma_{g}$ . Schematically this means that
\begin{equation}\label{cond}
    \gamma_{d}(R, t)=\left\{\begin{array}{ll}\Omega_{p}(R) t & \text { if } \mathrm{St}>>1 \\ \omega_{p} t & \text { if } \mathrm{St}<<1.\end{array}\right.
\end{equation}
In order to treat intermediate particle sizes, we make the ansatz that the twist profile of the dust can be written as a linear combination of the asymptotic behaviours
\begin{equation}
    \gamma_{d}(R, t, \mathrm{St})=A(\mathrm{St}) \Omega_{p}(R) t+B(\mathrm{St}) \omega_{p} t,
\end{equation}
where $A,B$ are functions of the Stokes number. These two functions have some constraints to respect: (i) it is necessary to recover the conditions (\ref{cond}) for large and small Stokes number and (ii) the position of the co-precession radius should not be modified, meaning that
\begin{equation}
    A=1-B.
\end{equation}
One possible choice for the functions is the following
\begin{equation}
    A=\frac{\text{St}}{1+\text{St}},\quad B=\frac{\text{1}}{1+\text{St}},
\end{equation}
which gives
\begin{equation}
    \gamma_{d}(R, t, \mathrm{St})=\frac{\mathrm{St}}{1+\mathrm{St}} \Omega_{p}(R)+\frac{1}{1+\mathrm{St}} \omega_{p}.
\end{equation}
For small Stokes number, the dust precession profile becomes shallower and $\Delta\omega\to0$: this will affect the position of dust rings because 
\begin{equation}
    \lim_{\Delta\omega\to0}\cos\psi = 1,
\end{equation}
meaning that, for small Stokes number, the formation of these structures is more difficult. Indeed, the dust is well-coupled to the gas and also precesses together with it. In particular, for decreasing $\text{St}$, the position of $R_\text{IDR}$ and $R_\text{ODR}$ goes outside the disc, as can be seen in figure \ref{ringsStokes}, implying no ring formation. In practice, we expect ring formation only for $\text{St}\gtrsim 1$.

\section{Numerical setup}\label{S3}
We have performed a suite of numerical simulation using the SPH code PHANTOM \citep{PHANTOM}. This code can be used to study the evolution of dust and gas considered as a single fluid -with the so called one fluid implementation, suitable for values of the stokes number $\text{St} < 1$ \citep{1fluid1,1fluid2}- or as a set of different fluid -the two fluids method, for fluids with $\text{St} >1$ \citep{2fluid1,2fluid2}. This code is widely used in astrophysical community to study gas and dust dynamics in accretion discs, both using one fluid \citep{1f3,1f2,1f1} and two fluids \citep{hltau} methods.

In this work, we decided to use the two fluids method because we are interested in the marginally coupled regime ($\text{St}\gtrsim1$). In this method, the code solves two sets of hydrodynamical and a drag term that takes account for the aerodynamical coupling.

\begin{table}
\begin{tabular}{ccccccc}
$\#$ & $M_{1} / M_{2}$ &  $R_{\text {in }}[\mathrm{AU}]$ & $R_{\text {out }}[\mathrm{AU}]$ & $p$ & $q$ & $H / R$  \\
\hline S1 & 1 & 15 & 150 & $3 / 4$ & $3 / 4$ & 0.1  \\
S2 & 1 &  15 & 150 & $5 / 4$ & $3 / 4$ & 0.1  \\
S3 & 1 &  15 & 150 & 1 & $3 / 4$ & 0.1 \\
S4 & 1 &  15 & 100 & 1 & $3 / 4$ & 0.1 \\
S5 & 1 &  15 & 75 & 1 & $3 / 4$ & 0.1 \\
S6 & 2 &  15 & 150 & 1 & $3 / 4$ & 0.1 \\
S7 & 1 &  15 & 100 & 1 & $1 / 2$ & 0.1 \\
S8 & 1 &  15 & 150 & $3 / 4$ & $3 / 4$ & 0.125\\
\hline
\end{tabular}\caption{Parameters set for the simulations. Note that we changed only values that determines the position of the dust rings.}\label{table1}
\end{table}

\begin{table}
\begin{tabular}{cccccc}
$\#$ & $\rho_{0}\left[\mathrm{g} / \mathrm{cm}^{3}\right]$ & $s[\mathrm{cm}]$ & $\Sigma_{d}^{\mathrm{in}}\left[\mathrm{g} / \mathrm{cm}^{2}\right]$ & Approx. St & $f$ \\
\hline $\mathrm{S} 1$ & 5 & 15 & 0.093 & 15 & 0.01 \\
$\mathrm{S} 2$ & 5 & 10 & 0.299 & 5& 0.01 \\
$\mathrm{S} 3$ & 5 & 20 & 0.135 & 15 & 0.01\\
$\mathrm{S} 4$ & 5 & 15 & 0.221 & 10 & 0.01\\
$\mathrm{S} 5$ & 5 & 15 & 0.222 & 5 & 0.01\\
$\mathrm{S} 6$ & 5 & 20 & 0.134 & 15 & 0.01\\
$\mathrm{S} 7$ & 5 & 25 & 0.365 & 10 & 0.01\\
$\mathrm{S} 8$ & 5 & 15 & 0.093 & 15 & 0.01\\
\hline
\end{tabular}\caption{Properties of dust in the simulations performed in this work.}\label{table2}
\end{table}

We performed a set of 8 simulations trying to explore as deeply as possible the parameter space (see table \ref{table1}), all with grains marginally coupled $\text{St}\gtrsim1$ (see table \ref{table2}). The other parameters of the simulations have been chosen as follows: the semi major axis of the binary stellar system $a=\SI{10}{AU}$, their eccentricity $e=0$, the inner radius of the circumbinary disc $R_\text{in} = \SI{15}{AU}$, the accretion radius of the two stars $R_\text{acc}=\SI{5}{AU}$, the inclination of the circumbinary disc $\beta = \pi/6$, the total (gas and dust) mass of $\SI{0.02}{M_\odot}$, with a dust-to-gas ratio $f=1/100$ and the $\alpha-$viscosity coefficient \citep{alphaSS} is taken to be $\alpha = 0.01$. All our simulated systems are gravitationally stable, therefore no self gravity is computed. We choose this set of parameters in analogy with \citet{AlyLod}, in order to confirm their results and to expand them. 

In all our simulations we included $N_g=10^6$ gaseous particles, initially distributed with an exponentially tapered surface density profile
\begin{equation}
    \Sigma_{g}(R, t=0)=\Sigma_{\mathrm{g}, \mathrm{0}}\left(\frac{R}{R_{\mathrm{in}}}\right)^{-p} \exp \left[-\left(\frac{R}{R_{\mathrm{out}}}\right)^{2-p}\right],
\end{equation}\label{density0}
where we refer all quantities to the internal radius $R_\text{in}=\SI{15}{AU}$. We adopt a locally isothermal equation of state $P=c_s^2\rho_g$ with
\begin{equation}
    c_{s}(R)=c_{s, \text { in }}\left(\frac{R}{R_{\text {in }}}\right)^{-q},
\end{equation}
and thus the thickness of the disc can be written as
\begin{equation}
    H(R)=H_{\mathrm{in}}\left(\frac{R}{R_{\mathrm{in}}}\right)^{-q+3 / 2}.
\end{equation}

As for the dust disc, the number of dust particles for every simulation is $N_d = 10^5$ initially distributed with the same density profile of the gas. All numerical simulations are performed with only one kind of grain size with intrinsic dust density $\rho_0 = \SI{5}{g/cm^3}$, and size spanning from $10$ to $\SI{25}{cm}$: these properties assure that the drag force regime is Epstein's one. The algorithm computes the Stokes number as
\begin{equation}
    \mathrm{St}=\frac{\rho_{0} s}{\rho c_{\mathrm{s}}} \sqrt{\frac{\pi \gamma}{8}} \Omega,
\end{equation}
where $\rho=\rho_g +\rho_d$, and its computational time is long for $\text{St}\gtrsim1$. For computational convenience, we used the prescription of \citet{poblete}: we dropped $\rho_d$ in the sum and we used $\rho=\rho_g$ in the code. By doing so, the stopping time is overestimated by a factor of $1+f$, with $f<1$ the dust-to-gas ratio, and therefore we obtained significant results for the dust evolution in shorter computational time.

\section{Numerical results}\label{S4}

\subsection{Co-precession radius}
\begin{figure}
    \hspace{0.15cm}
    \includegraphics[scale=0.455]{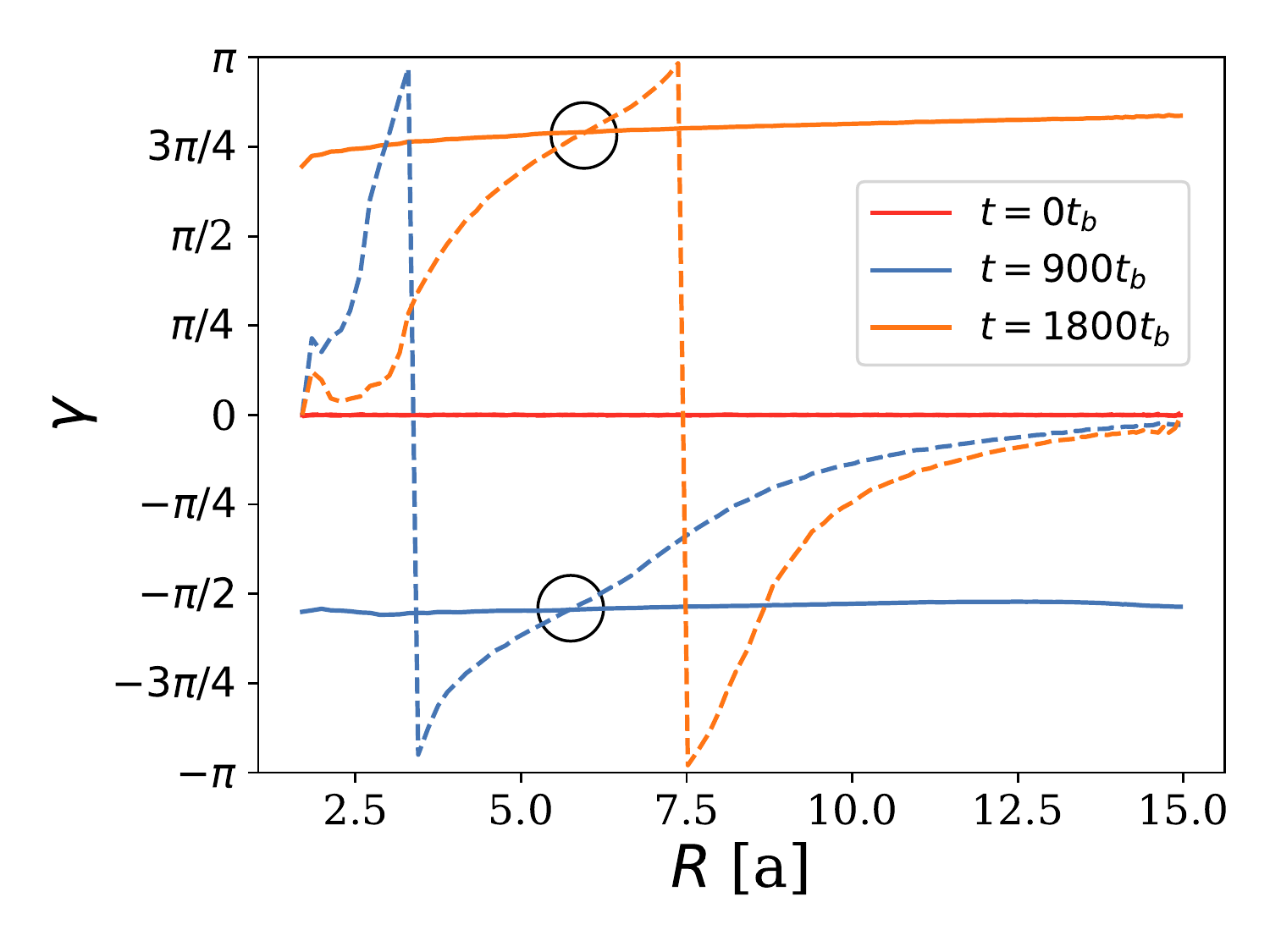}
    \caption{Twist profiles of gas (solid line) and dust (dashed line) for S1 at different times. The circles show the radius at which they meet, i.e. the co-precession radius.}
    \label{twists1}
\end{figure}

In order to estimate the co-precession radius from numerical simulations, we need to compare the twist profile of gas and dust. As can be clearly seen in figure \ref{twists1}, that shows the twist angle $\gamma$ as a function of time, in units of the binary period $t_b$, the two components precess at different rates. On one hand, the gaseous disc, that communicates viscously, precesses rigidly; on the other hand, the dust, being a pressureless and viscousless fluid, differentially precesses. The two components are out of phase almost everywhere, except for one point, the co-precession radius.

\begin{figure}
    \centering
    \includegraphics[scale=0.42]{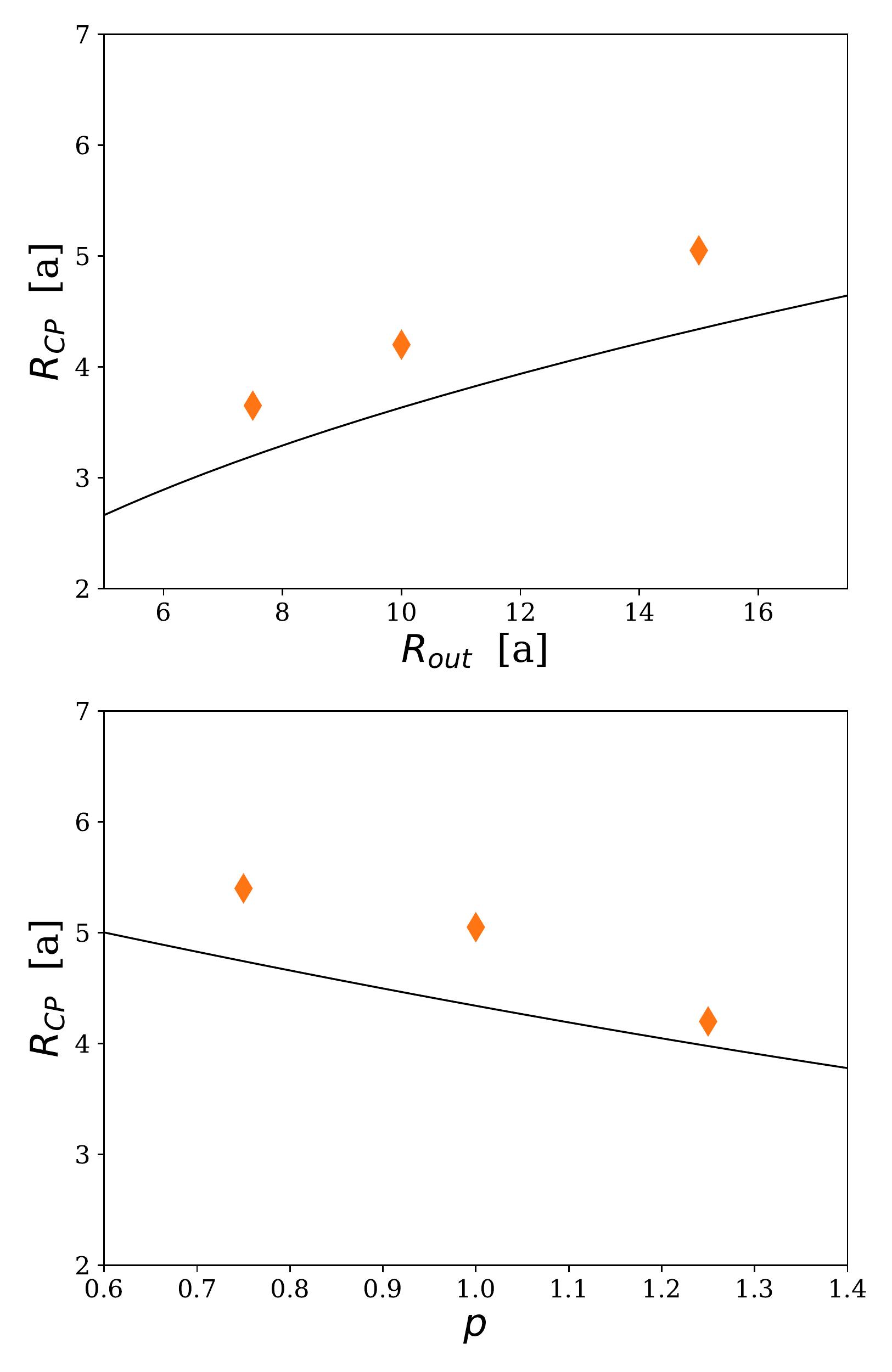}
    \caption{Comparison between the theoretical expectations (black lines) and the simulations' results (red diamonds) for the position of the co-precession radius for simulations with different outer radius R$_{\rm out}$ and $p$ index.}
    \label{results_cp}
\end{figure}

\begin{figure}
    \centering
    \includegraphics[scale=0.455]{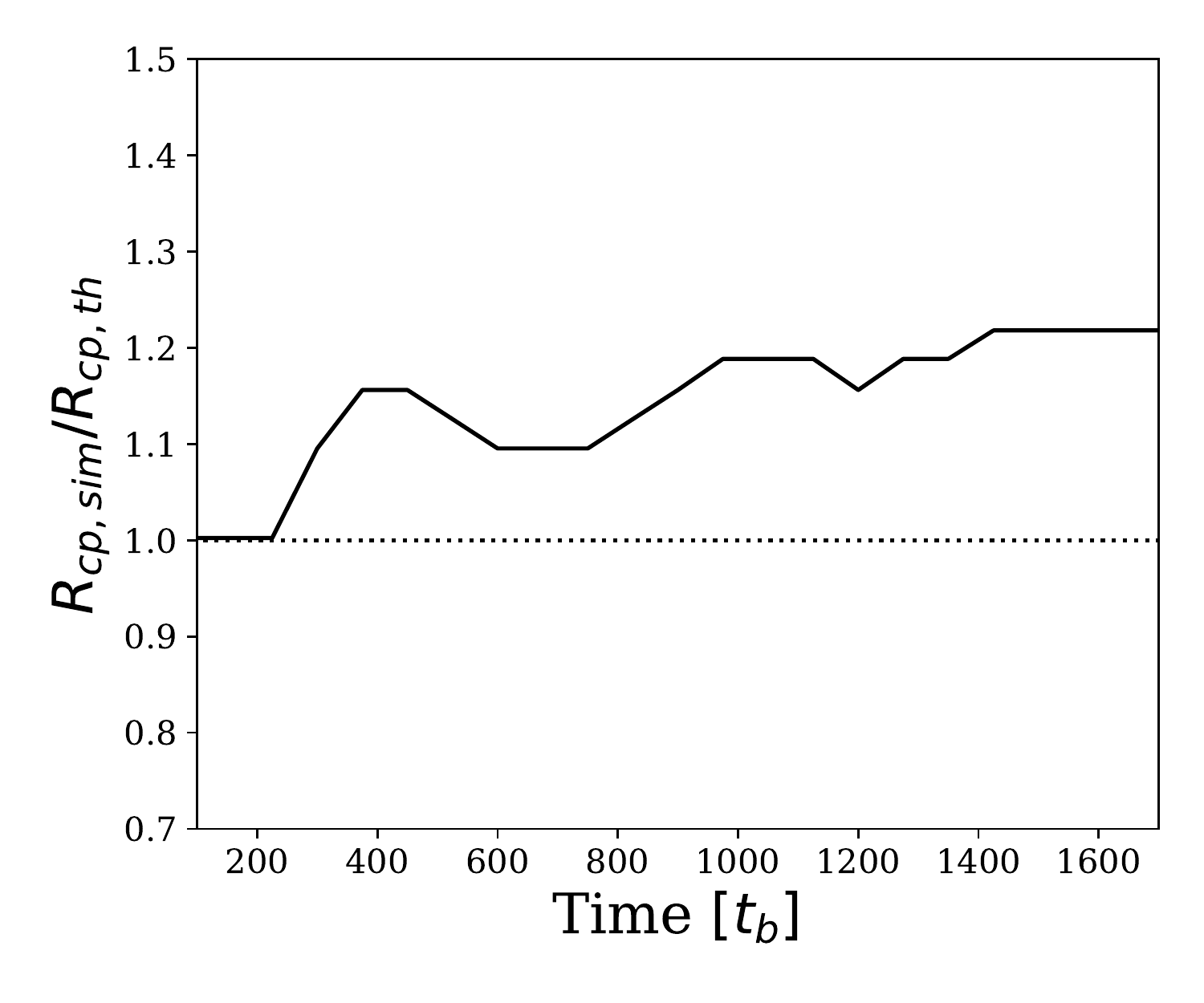}
    \caption{Time evolution of the position of the co-precession radius in simulation S1 normalised over the theoretical one. Note that for $t\to 0$ the result of the simulation is coincident with the theoretical expectation. The evolution of the density profile introduces an error of about $25$\% over the course of the simulation.}
    \label{evolution}
\end{figure}

Figure \ref{results_cp} shows the comparison between the theoretical expectations and the numerical results for the co-precession radius. Although the simulations give a higher value of the co-precession radius (discrepancy of $\sim25\%$), the trend of the curve is respected. The discrepancies can be explained as a consequence of the gaseous surface density evolution. Indeed, the co-precession radius is sensitive to the shape of the gaseous density profile, because it determines its precession frequency $\omega_p$: in our theoretical model, we have assumed that the shape of the density profile is time independent. However it evolves, and so does the position of $R_\text{cp}$, as can be seen in figure \ref{evolution}: for increasing time, the co-precession radius shows an increasing discrepancy with respect to the theoretical prediction. In any case, for $t\to 0$ the two values are in agreement.

\subsection{Tilt profile: alignment and lost of coplanarity}\label{alignmentsection}
\begin{figure}
    \hspace{0.15cm}
    \includegraphics[scale=0.455]{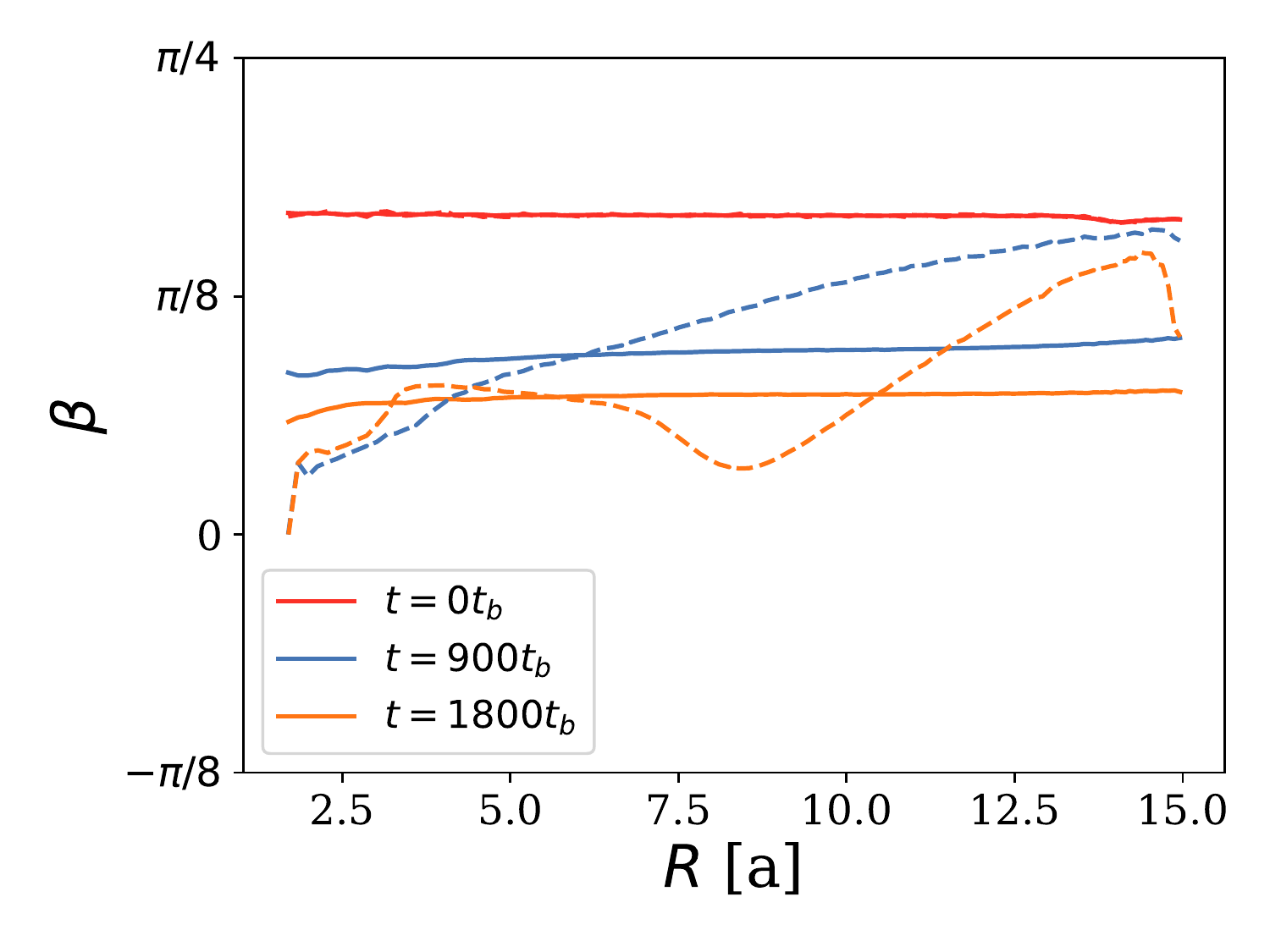}
    \caption{Tilt profiles of gas (solid line) and dust (dashed line) of S1 for different times.}
    \label{tilts1}
\end{figure}

According to our ansatz for the analytical estimate, the tilt angles of dust and gas are the same and do not vary with time (see section \ref{velocityfields}). We already know that it is not true: indeed, the gas disc tends to align with the binary plane with a timescale $t_{a}$ of the order of the viscous time \citep{align1, align2,align3}. In addition, in figure \ref{tilts1} it can be clearly seen that the gaseous disc tends to align to the binary plane $(\beta=0)$. The dust disc has also a peculiar behaviour, showing a variation in the tilt profile between small and large radii.

How can we explain the tilt profile of the dust? In general, the dust is influenced by both aerodynamical drag and gravitational force. As already stated, the strength of the aerodynamical coupling depends on the Stokes number: uncoupled particles mainly experience the gravitational force, that makes the dust vertically oscillate around the plane $z=0$; on the other hand, tightly coupled dust particles tend to follow the motion of the gas disc, that is the alignment with the binary plane within a timescale $t_a$. Thus, the role of the drag force is to damp the vertical oscillations of the dust, with the alignment time as a typical timescale. 

In the work \citet{AlyLod}, the authors performed a series of simulations of a misaligned circumbinary disc. They tried to reproduce different coupling between gas and dust ($\text{St}\in[0.002,1000]$) and they saw that for large Stokes number the tilt profile of the dust shows oscillations that gradually transform the dusty disc into a spherical cloud, while for small $\text{St}$ the oscillations damp.

\begin{figure*}
    \centering
    \includegraphics[scale=0.46]{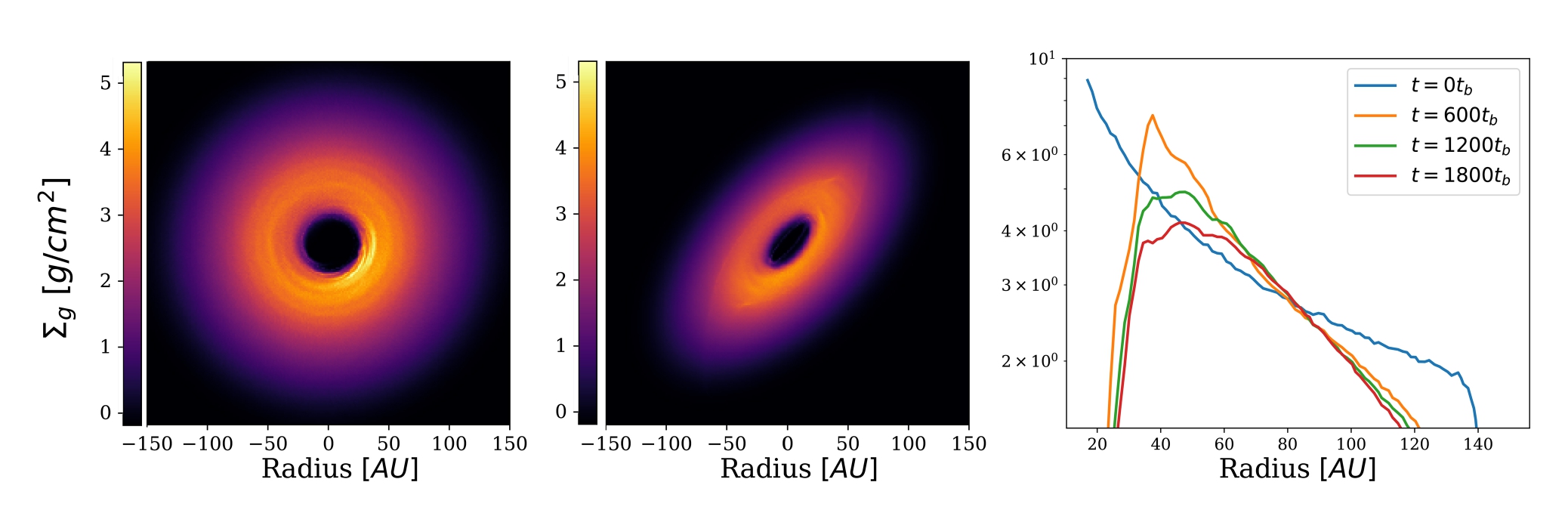}
    \includegraphics[scale=0.46]{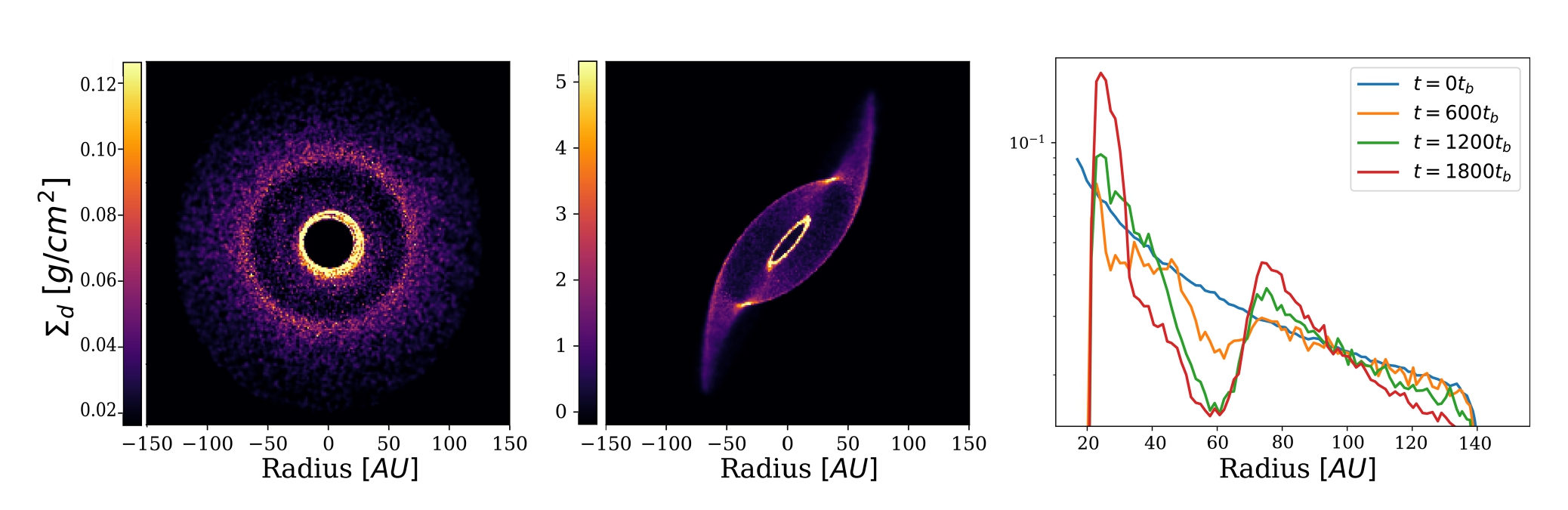}
    \caption{Snapshots of simulation S1 showing the density of both gas (top panel) and dust (bottom panel). On the left and centre: surface density of gas (top) and dust (bottom) from different perspectives ($i=0$ and $i=2\pi/5$) for $t =1800t_b =\SI {63 000} {yr}$. On the right: comparison between gas (top) and dust (bottom) surface density profile for different times. Clearly, the dust trap mechanism is not caused by the gas, but it is self-induced in the dusty component. Indeed, the gas density profile does not show any peculiar signature, while in the dusty one two rings are clearly visible. }
    \label{snaps}
\end{figure*}
\begin{figure*}
    \centering
    \includegraphics[scale=0.38]{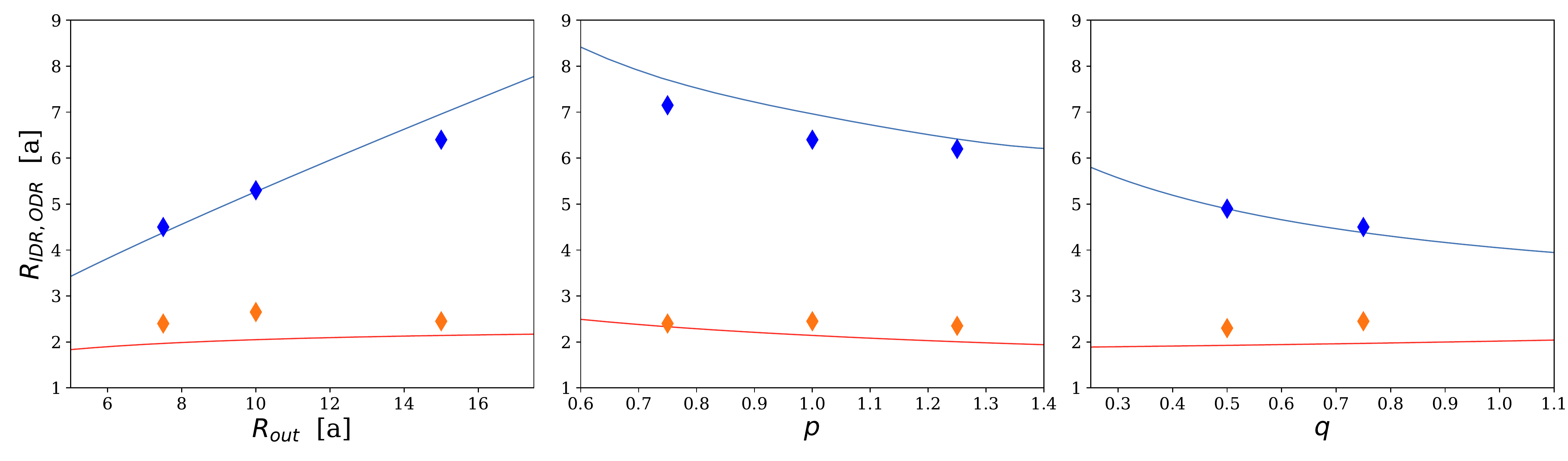}
    \includegraphics[scale=0.38]{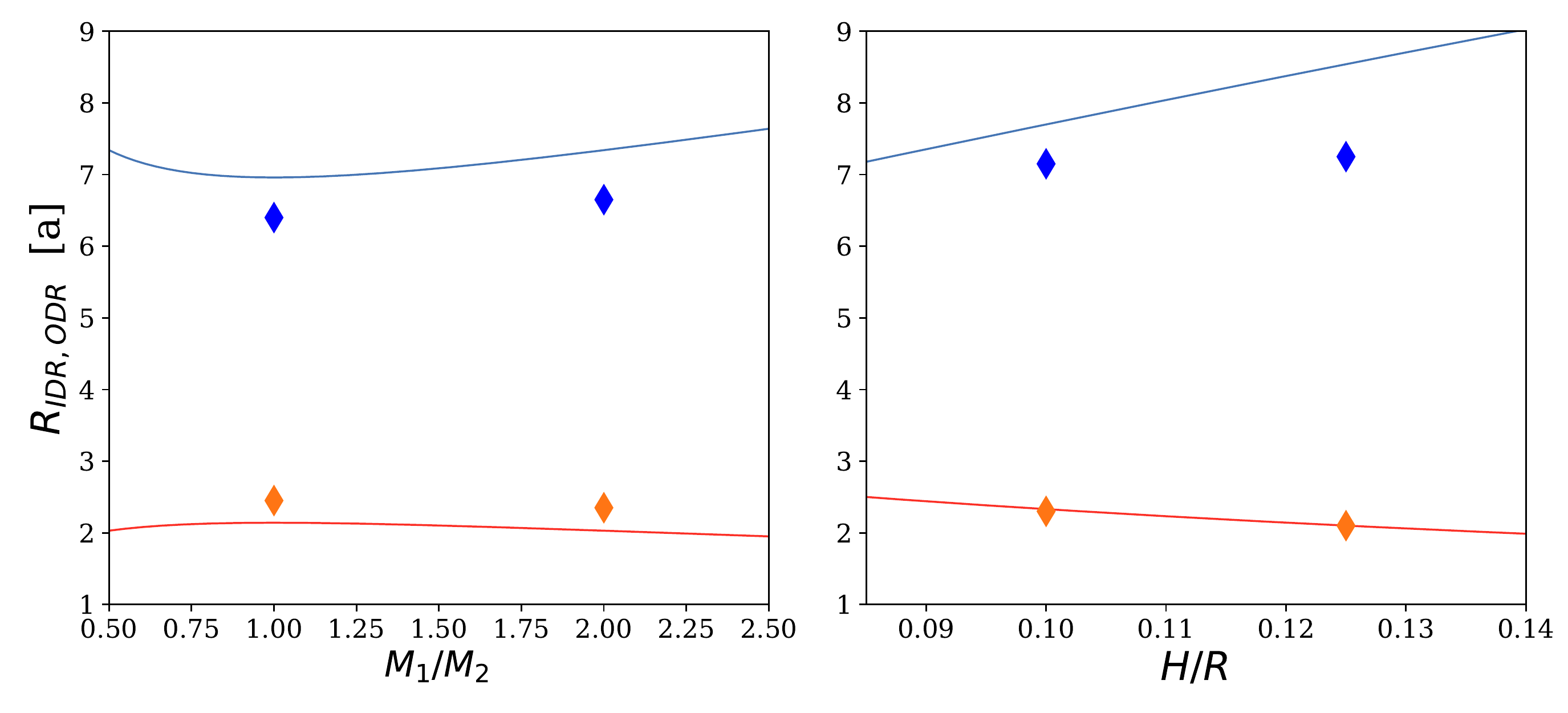}
    \caption{Comparison between the theoretical expectations (lines) and the simulations' results (dots). The solid lines represents the theoretical model with an exponential tapered density profile and the dots are the results of the simulations. The red objects refers to $R_\text{IDR}$ while the blue ones to the $R_\text{ODR}$.}
    \label{resultz}
\end{figure*}

\subsection{Dust rings}

To find the position of the rings, we analyse the dust surface density. This quantity is calculated at every spherical radius by averaging azimuthally with respect to the local direction of the angular momentum. Figure \ref{snaps} shows a comparison between the surface density profile of gas and dust. From these plots it is clear that there are two maxima in the dust density profile (i.e. the two rings) while no features appear in the gas component. This means that these pile ups do not form in any gas pressure maxima: they are non-conventional dust traps.

Regarding the position of dust rings, figure \ref{resultz} shows the comparison between the results of the simulations and the theoretical expectations. Usually the inner dusty ring forms very close to the inner radius $R_\text{in}$. In this region we have to take into account second order effects due to the binary system, as density waves and tidal forces. For that reason, the position of the inner ring found in the simulations is larger than the one predicted by the theoretical model: tidal forces in the inner part of the disc tend to push outwards the ring. The outer ring shows a better agreement between the simulations and the theoretical model. It has also another important property: all the dust coming from the outer part of the disc migrates towards inner radii because of radial drift and, when it meets the outer dusty radius, it piles up. This leads the outer trap to accumulate more dust, significantly enhancing its mass, since the outer part of the disc contributes to its growth.

\section{Connection with observations}\label{S5}
\begin{figure*}
    \centering
    \includegraphics[scale=0.41]{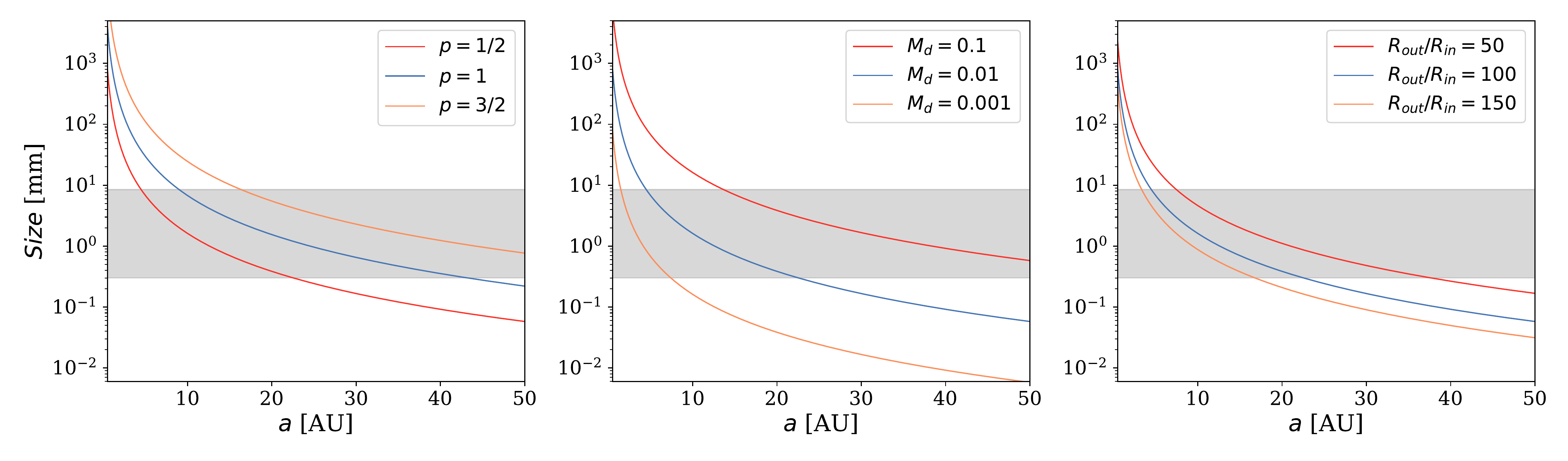}
    \includegraphics[scale=0.41]{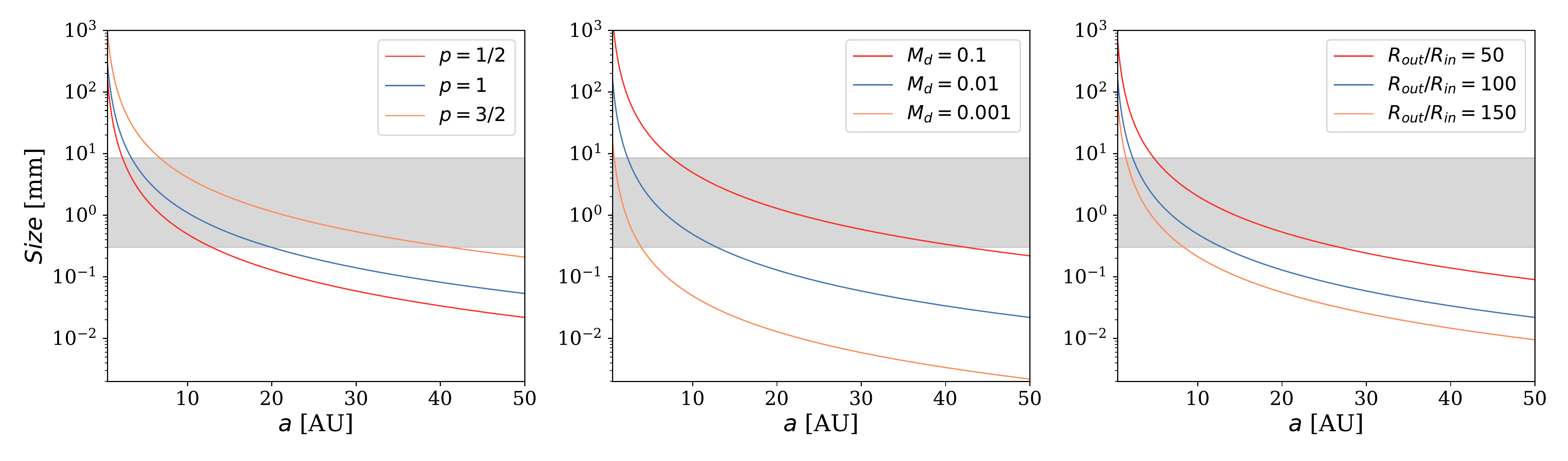}
    \caption{Critical size of dust particle as a function of binary separation evaluated at the inner (top) and outer (bottom) ring. First column: size for different values of the density parameter $p$, with $M_\text{d} = 0.01\text{M}_*$, $x_\text{out}=100$. Second column: size for different values of the disc mass $M_d$, with $x_\text{out}=100$ and $p=1/2$. Third column: size for different values of $x_\text{out}$, with $M_\text{d} = 0.01\text{M}_*$ and $p=1/2$. The grey part of the plots refers to the observational window of ALMA [0.3mm; 8.5mm].}
    \label{sizein}
\end{figure*}

So far, we have shown that dust rings are likely to form for marginally coupled dust particles $(\text{St}\gtrsim1)$. In this section, we present a simple analysis to show how to connect this critical Stokes number regime with particle size and binary separation, in order to study the possible observability of these structures.

As we have already stated, the effect of the radial drift is larger for 
\begin{equation}
    \mathrm{St}=\frac{\pi \rho_{0} s}{2 \Sigma_{g}} \simeq 1,
\end{equation}
which gives an expression for the size ath which $\text{St}= 1$ depending on disc parameters
\begin{equation}\label{grainsize}
    s(R)=\frac{2}{\pi} \frac{\Sigma_{g}(R)}{\rho_{0}}=\frac{2}{\pi \rho_{0}} \frac{M_{d}(2-p)}{2 \pi R_{\mathrm{in}}^{2}} \frac{1}{x_{\mathrm{out}}^{2-p}-1}\left(\frac{R}{R_{\mathrm{in}}}\right)^{-p},
\end{equation}
where we have assumed a pure power law density profile for the gas. With equation (\ref{grainsize}) we are able to evaluate the critical size of dust particles at any radius $R$ depending on disc parameters. In figure \ref{sizein} we show the size of dust particles with $\text{St}=1$ at inner and outer rings as a function of the binary separation for different disc masses, power law index of the density profile and disc dimension. The grey rectangle is the observational window of ALMA telescope: for a wide range of parameters, the dust rings are visible with this interferometer. The interesting feature of a system like this is that the rings are present only in the dusty component, thus, in principle, observing the CO no structures would be visible.

A crucial thing to note is that dust rings occur at any binary separation, and the problem is completely scalable. Indeed, the semi-major axis determines the size of dust grains with St=1, and thus the ALMA observability.

\subsection{The case of GW Orionis}
In this paragraph we apply what we have found in a real system. Although no misaligned circumbinary disc with two dust dust rings has been observed, we consider a circumtriple disc that, under certain approximations, may suit our purposes.

Recent observations of the triple star system GW Orionis have shown a peculiar geometry, consisting of a circumtriple broken misaligned disc with three dust rings \citep{gwori1,gwori2}. GW Orionis is a young hierarchical triple stellar system (${1\pm0.1}\text{Myr}$, \citet{ageOri}) in the Orion Molecular Cloud at a distance of $388\pm5\text{pc}$ \citep{oridist}. It consists of a close binary system (GW Ori A and GW Ori B) with masses $M_A = 2.47\pm0.43\text{M}_\odot$, $M_B=1.43\pm0.18\text{M}_\odot$ and semi-major axis $a_1 = \SI{1.2}{AU}$ and a third star (GW Ori C) with mass $M_C=1.36\pm0.28\text{M}_\odot$ that orbits with a distance from the centre of mass of the AB system $a_2 = \SI{8.5}{AU}$ \citep{gwori2}. The circumtriple disc is broken very close to the stars: the inner part shows a dust ring aligned to the orbital plane of AB-C while the outer disc shows two dust rings, misaligned to the orbital plane AB-C by an angle $\beta\sim40^\circ$. The inner ring has a position $R_0=\SI{43}{AU}$, the outer ones $R_1=\SI{182}{AU}$ and $R_2=\SI{334}{AU}$.

This system has a non-keplerian potential and a misalignment, that make the formation of dust rings possible. Now we want to see if there is a reasonable combination of parameters that reproduce the observed position of the outer ring. At large distances, the triple system can be approximated as a binary, with $M_1=M_A+M_B$ and $M_2=M_C$ and semi-major axis $a=a_2$. Fixing $R_\text{in}=5a$, $R_\text{out}=52a$ \citep{gwori2}, we performed an optimisation of the parameters and we found that the position of the outer ring is better reproduced with $p=1.75$, $q=0.85$ and $(H/R)_\text{in}=0.05$.  With these parameters, the expected position is $R_2^\text{exp}=\SI{357}{AU}$. We tried to reproduce only the external ring because the inner one is too close to the breaking and thus our theoretical construction does not work. In addition, in the inner part of the system, the binary approximation is not valid enough, because effects of the triplicity arise.

\section{Conclusions}\label{S6}
In this work we analytically and numerically studied the problem of dust traps formation in misaligned circumbinary discs. We found that these pile-ups may be induced not by pressure maxima, as the usual dust traps, but by a dynamical mechanism. This consists in the formation of two regions in the disc where the velocity difference between the two components is zero, thus no radial drift could occur. The position of these rings depends on a series of parameters defining the disc, as its radial extent, its density profile and its thickness. At these locations, the dust, that migrates from larger radii because of radial drift, piles up and this leads to the formation of dusty rings. This analysis holds for high Stokes numbers, when the precession profile of the dust is not influenced by the gas. For small Stokes numbers, the dust precession profile becomes shallower because of the drag force. This effect makes the formation of dust rings more difficult

To confirm our predictions, we then performed a series of numerical simulations using the two-fluids mode of the SPH code PHANTOM. By means of these simulations we explored the parameter space as much as possible: we found an overall general agreement between the results of the simulations and the analytical expectations.

This mechanism is important in the context of planet formation because it fosters dust growth, piling up solid particles in the two rings. Thus, in principle, these are the locations in which planetary formation can take place. In addition, these rings are theoretically visible with ALMA telescope, because the critical size of dust particles is compatible with the observational window of the interferometer. We stress that the peculiar feature of a system like this is that rings are present only in the large dust: gas observation (CO isotopologues or scattered light) would not show any particular structure.

The two main requirements to form dust rings with this mechanism are the lack of keplerian potential (in this case, due to the binary stellar system) and the misalignment (that allows the precession): in principle, this mechanism might explain the formation of dust rings not only in circumbinary systems, but also in multiple-star discs with a misalignment, as we have discussed for GW Orionis.

\section*{Acknowledgements}
The authors thank the referee for useful suggestions, in particular about the resolution test. This project and the authors have received funding from the European Union's Horizon 2020 research and innovation programme under the Marie Skłodowska-Curie grant agreement No 823823 (DUSTBUSTERS RISE project). Computational resources have been provided by INDACO Platform, a project of High Performance Computing at the Università degli Studi di Milano. CL and CT thank Nicolás Cuello, Benedetta Veronesi, Simone Ceppi and Alessia Annie Rota for useful discussions. This work made use of the Python packages Numpy and matplotlib and of Plonk \citep{plonk}, a smoothed particle hydrodynamics analysis and visualisation with Python. HA acknowledges funding from ANR (Agence Nationale de la Recherche) of France under contract number ANR-16- CE31-0013 (Planet-Forming-Disks) and thank the LABEX Lyon Institute of Origins (ANR-10-LABX-0066) of the Université de Lyon for its financial support within the programme ‘Investissements d’Avenir’ (ANR-11-IDEX-0007) of the French government operated by the ANR.
\section*{Data Availability}
The data presented in this article will be shared on reasonable request to the corresponding author.
 



\bibliographystyle{mnras}
\bibliography{bibliography} 

\appendix

\section{Rigid body precession}\label{appA}
In this appendix we estimate the global gaseous precession frequency $\omega_p$ using the same procedure as in \citet{Lodato2014}. At a given radius $R$ within the disc, the local external torque density is $\mathbf{T}(R)=\mathbf{\Omega}_{{p}}(R) \times \mathbf{L}(R)$ \citep{lodpr}, where $\mathbf{\Omega}_{p}(R)$ is the free precession induced by the binary system $(\Omega_p\propto R^{-7/2})$ and $\mathbf{L}(R)=\Sigma \Omega R^{2} \Bell$ is the angular momentum per unit area of the disc. We define the vector $\mathbf{\Omega}_p(R),$ to be along the $z$ -direction. If the warp amplitude is small, we can write the scalar equation for the local external torque density
\begin{equation}
    T(R)=\Omega_{p}(R) L(R),
\end{equation}
where $L(R)=\Sigma R^{2} \Omega \sqrt{\ell_{x}^{2}+\ell_{y}^{2}}$ is the angular momentum projected on the $xy$ plane. If we assume that the gaseous disc precesses rigidly with the characteristic frequency $\omega_p$, we can write
\begin{equation}
    T_\text{tot} = \omega_p L_\text{tot},
\end{equation}
where we have defined the two quantities
\begin{equation}
    T_{\text{tot}}=\int_{R_{\mathrm{in}}}^{R_{\mathrm{out}}} \Omega_p(R) L(R) 2 \pi R \mathrm{d} R,
\end{equation}
and
\begin{equation}
    L_{\text{tot}}=\int_{R_{\mathrm{in}}}^{R_{\mathrm{out}}} L(R) 2 \pi R \mathrm{d} R.
\end{equation}
These expressions are appropriate for small warps, that is the case we are considering. Hence the global precession frequency can be written as
\begin{equation}
    \omega_{{p}}=\frac{\int_{R_{\mathrm{in}}}^{R_{\mathrm{out}}} \Omega_{{p}}(R) L(R) 2 \pi R \mathrm{d} R}{\int_{R_{\mathrm{in}}}^{R_{\mathrm{out}}} L(R) 2 \pi R \mathrm{d} R} = \Omega_p(R_\text{in})\xi, 
\end{equation}
where $\xi$ depends exclusively on the gaseous surface density profile $\Sigma$. In order to compute $\xi$, we have to solve the following integral
\begin{equation}
    \xi = \frac{\int_{1}^{x_{\text {out }}} x^{-2}\sigma(x) \text{d} x}{\int_{1}^{x_{\text {out }}} x^{3 / 2} \sigma(x) \text{d} x} ,
\end{equation}
where $\sigma(x)$ is the adimensional surface density profile $\sigma=\Sigma/\Sigma_\text{in}$. In \citet{Lodato2014}, $\xi$ is computed for a pure power law density profile $\sigma(x) = x^{-p}$, giving
\begin{equation}
    \xi_\text{pl} = \frac{\int_{1}^{x_{\text {out }}} x^{-2-p} d x}{\int_{1}^{x_{\text {out }}} x^{3 / 2-p} d x} = \frac{1-x_{\text {out }}^{-1-p}}{x_{\text {out }}^{5 / 2-p}-1} \frac{5 / 2-p}{1+p};
\end{equation}
here we make the same calculation for an exponentially tapered density profile $\sigma(x) = x^{-p}\exp[-(x/x_\text{out})^{2-p}]$, that is
\begin{equation}
    \xi_{\exp }=  \frac{\int_{1}^{x_{\mathrm{out}}} \exp \left[-\left(\frac{x} { x_{\mathrm{out}}}\right)^{2-p}\right] x^{-2-p}\mathrm{d} x}{\int_{1}^{x_{\mathrm{out}}}  \exp \left[-\left(\frac{x} { x_{\mathrm{out}}}\right)^{2-p}\right] x^{3 / 2-p}\mathrm{d} x}  =x_{\mathrm{out}}^{-7 / 2} \frac{\Gamma_{\left[x_{\mathrm{out}}^{p-2}, 1\right]}\left(\frac{p+1}{p-2}\right)}{\Gamma_{\left[x_{\mathrm{out}}^{p-2}, 1\right]}\left(\frac{2 p-5}{2 p-4}\right)},
\end{equation}
where $\Gamma_{[a,b]}(y)$ is the generalised incomplete gamma function \citep{abramowitz}
\begin{equation}
    \Gamma_{[a, b]}(y)=\int_{a}^{b} \mathrm{d} t e^{-t} t^{y-1}.
\end{equation}

\section{Convergence test}\label{resolutiontest}
\begin{figure*}
    \centering
    \includegraphics[scale=0.35]{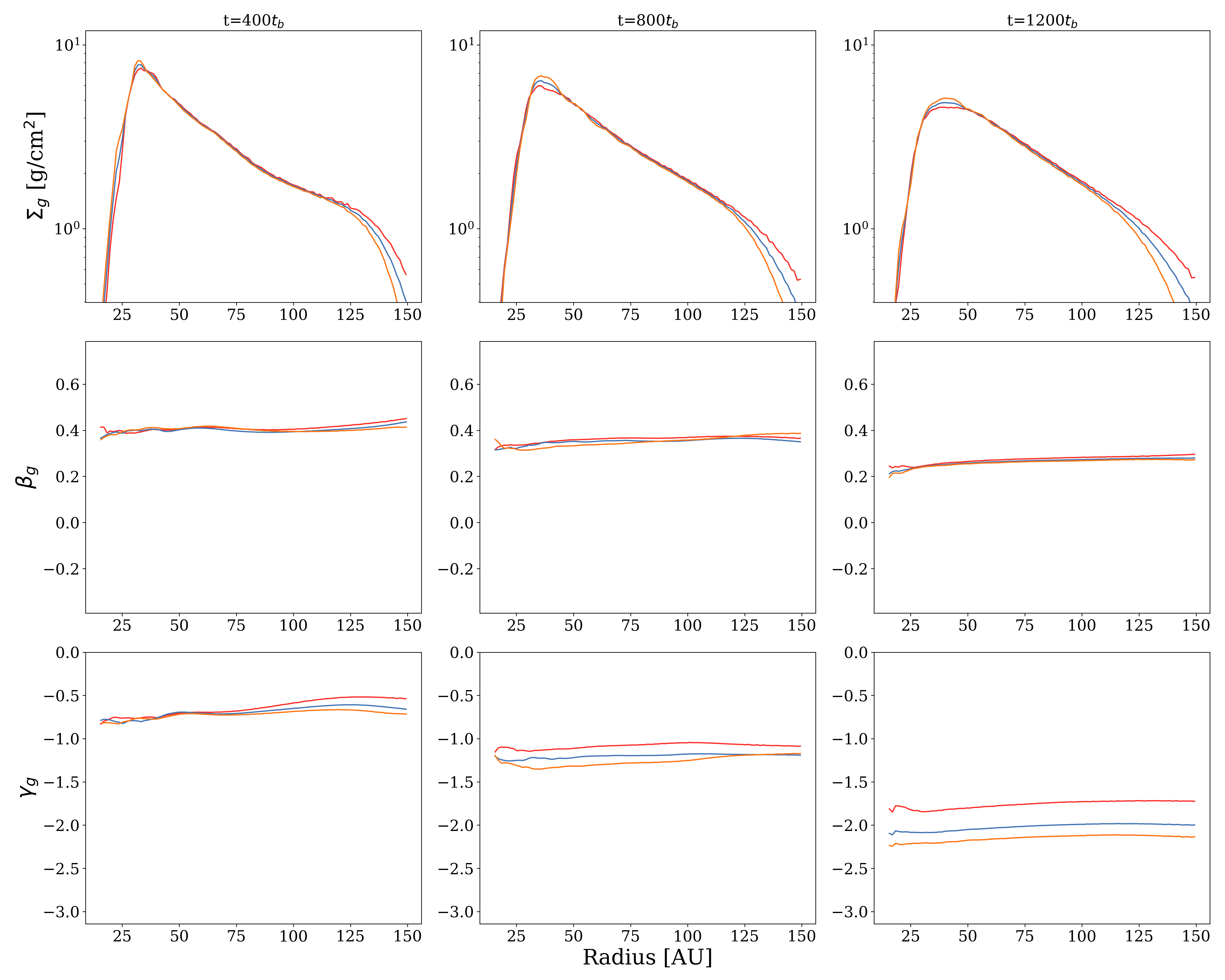}
    \caption{Resolution test for the gas component for the simulation S1. The red lines refers to the low resolution, the blue lines to the standard resolution and the orange lines to the high resolution. Top panel: comparison of the surface density profile for different times. Central panel: comparison of the tilt profiles. Bottom panel: comparison of the twist profiles.}
    \label{resgas}
\end{figure*}

\begin{figure*}
    \centering
    \includegraphics[scale=0.35]{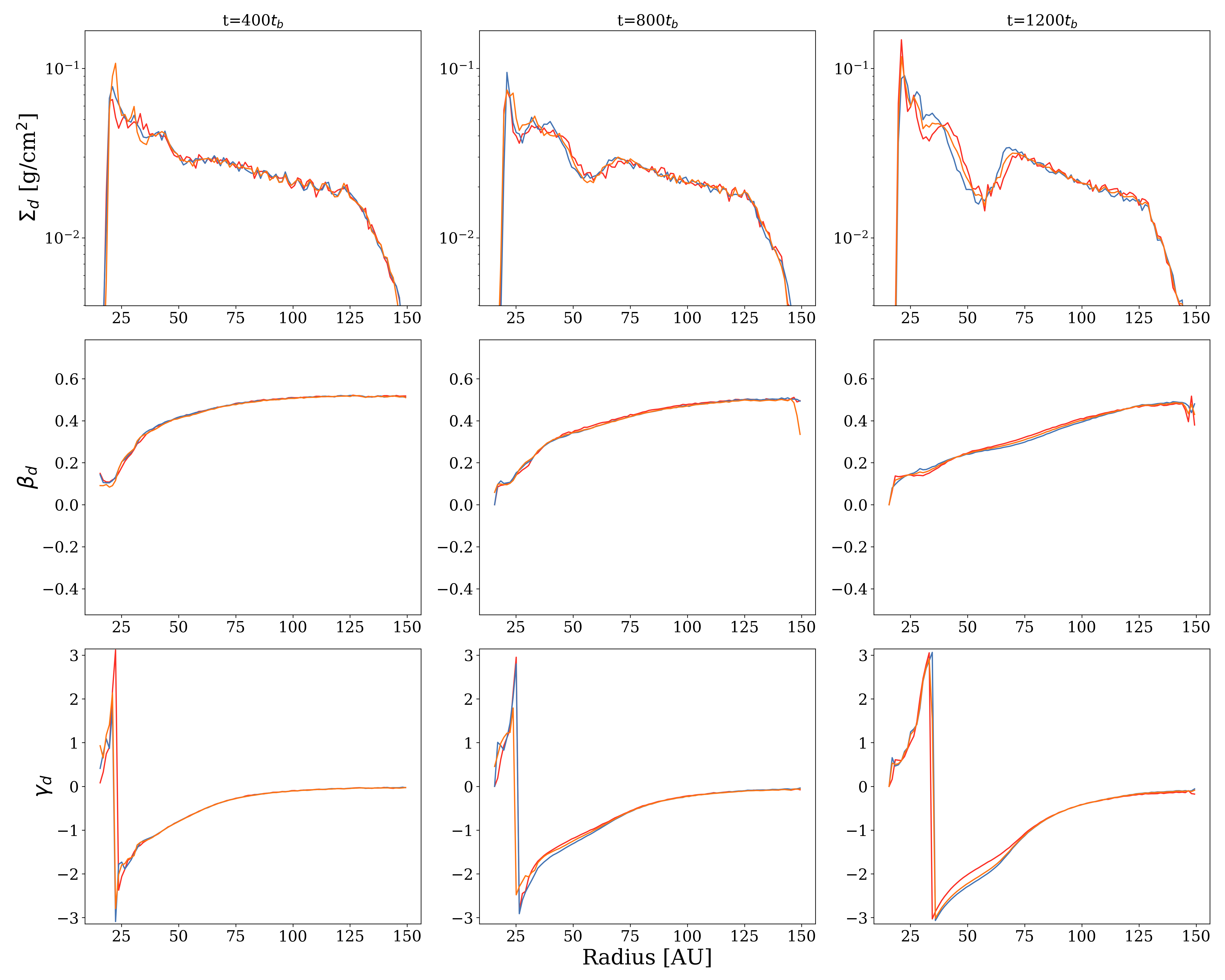}
    \caption{Resolution test for the dust component for the simulation S1. The red lines refers to the low resolution, the blue lines to the standard resolution and the orange lines to the high resolution. Top panel: comparison of the surface density profile for different times. Central panel: comparison of the tilt profiles. Bottom panel: comparison of the twist profiles.}
    \label{resdust}
\end{figure*}

In this appendix we present a convergence test of our simulations. To do this, we ran the simulation S1 (see table \ref{table1}) at half and twice the number of particles of the standard run, i.e. $N_g = 10^6$ and $N_d = 10^5$ particles. We do that in order to ensure that out results are robust. Figures \ref{resgas}, \ref{resdust} show the result of the convergence test: we compare the surface density, the tilt and twist profiles of both gas and dust for the three different resolutions. We can  say that our results are robust and resolution-independent: the profiles for different resolutions are in good agreement.

The gas evolution is also not significantly affected by the resolution. We do notice a slight difference in the surface density profiles due to increased viscous spreading for the lower resolution runs, which can be attributed to increased background dissipation intrinsic to particle reordering in SPH \citep{AV,dunhill}. There is also a slight difference in the gas disc precession speed, evident in the twist profiles on longer timescales, which can be attributed to better modelling of the disc thickness at higher resolution leading to a more accurate sound speed, which affects the wave propagation speed

\bsp	
\label{lastpage}
\end{document}